\documentclass[12pt]{article}
\usepackage{setspace}
\usepackage[margin=1in]{geometry}
\usepackage[utf8]{inputenc}
\usepackage[T1]{fontenc}
\usepackage{csquotes}
\usepackage{tikz}
\usetikzlibrary{positioning}
\usepackage[colorlinks=true, allcolors=blue]{hyperref}
\usepackage[round]{natbib}

\begin{document}

\title{Algorithmic Authority \& the Clinical Standard of Care}
\date{}

\author{
Aizierjiang Aiersilan \\
\small
The George Washington University \\
\small
alexandera@gwu.edu
}
\date{}

\maketitle

\begin{abstract}
The integration of artificial intelligence into clinical medicine creates a fundamental tension between algorithmic probabilistic reasoning and the experiential intuition of expert physicians; applying Lawrence Lessig's \enquote{Code is Law} framework, I argue that the architecture of clinical AI systems already functions as de facto medical regulation, reshaping liability and the standard of care. Reframing AI \enquote{hallucination} as structurally analogous to well-documented human cognitive failures such as confirmation bias and premature diagnostic closure, I show that both failure modes demand a unified governance response. I therefore propose a dialectical standard of care that treats the integrated AI-physician dyad as the singular responsible diagnostic entity, mandating the synthesis of algorithmic precision with human interpretive authority within robust data governance and patient privacy frameworks.
\end{abstract}

\section{Introduction}

The proliferation of AI in clinical medicine represents a profound shift in diagnostic epistemology. Foundation models capable of processing heterogeneous clinical data networks, including radiological imaging, structured electronic health records, genomic profiles, and free-text clinical notes, are being rapidly integrated into frontline medical settings~\citep{vela2025toward}. While current literature often characterizes this trajectory as a technical revolution where AI systems operate as primary diagnostic agents~\citep{lee2023ai}, this paradigm shift currently unfolds within a critical regulatory vacuum. Existing medico-legal frameworks were conceptualized for an environment where human physicians exclusively exercise independent diagnostic judgment~\citep{spinello2010cyberethics}.

I argue that the assimilation of AI into clinical decision-making surfaces a profound tension in how medical knowledge is produced and validated, what philosophers of science call an \emph{epistemological} tension. On one side stands the probabilistic reasoning inherent to machine learning systems; on the other, the tacit, experiential knowledge that constitutes clinical intuition. My perspective on this tension is informed by an interdisciplinary background spanning both computer science, where I have conducted research on AI-driven systems including in-context learning for autonomous agents~\citep{aiersilan2024generating} and AI-assisted body composition analysis for clinical screening~\citep{aiersilan2026biotoponet,aiersilan2026literature}, which together motivate the hybrid regulatory approach developed in this paper~\citep{aiersilan2026neuro}. Lawrence Lessig's foundational insight that \enquote{code is law}~\citep{lessig2009code} suggests that software architectures systematically regulate user behavior. As Lessig establishes, the codebase that structures a digital ecosystem serves as its constitution~\citep{lessig2009code}. This insight becomes acutely relevant in medicine, where algorithmic \enquote{code} generates diagnostic recommendations that explicitly shape the standard of care.

This paper proceeds in 5 distinct stages. First, I detail the two conflicting epistemologies: probabilistic machine reasoning and tacit clinical intuition. Second, I apply Lessig's framework to articulate how algorithmic design functions as de facto medical regulation. Third, I conceptually reframe the problem of AI \enquote{hallucination} as structurally parallel to cognitive failures in human clinical intuition. Fourth, I examine privacy and data governance as the necessary infrastructure for ethical AI integration. Finally, I elaborate on a proposed dialectical standard of care. This standard posits the AI-physician dyad as the locus of diagnostic responsibility, requiring a critical, structured review process that mitigates the risks of uncritical technological adoption. The central thesis is that navigating this epistemological tension requires an institutional architecture that treats the integrated physician-AI unit as the professional bearer of clinical responsibility. To build this argument, I first examine the 2 fundamentally distinct ways of knowing that collide when AI enters the clinic.

\section{Two Epistemologies in Tension}

\subsection{Probabilistic Machine Reasoning}

Contemporary clinical AI systems increasingly rely upon multimodal architectures capable of synthesizing diverse data streams. These systems integrate radiological imaging, structured EHR fields, and genomic markers into unified predictive frameworks~\citep{simon2025future}. The comparative performance of these models against human experts is well-documented. For instance, in the management of complex conditions such as aortic dissection, multimodal LLMs have achieved diagnostic accuracy matching that of multicenter panels of human clinical specialists~\citep{ekingen2026comparative}. Concurrently, benchmarking frameworks like SMMILE systematically evaluate multimodal medical in-context learning, establishing rigorous protocols to assess how foundation models adapt to novel clinical scenarios through few-shot exemplars~\citep{rieff2025smmile}.

The underlying epistemic mode of these systems is fundamentally probabilistic. They generate diagnostic outputs by computing conditional probability distributions over potential diseases or clinical interventions. This probabilistic foundation yields characteristic computational strengths, including high reproducibility and vast pattern recognition capabilities across high-dimensional data, alongside inherent vulnerabilities such as acute sensitivity to dataset distributional shifts.

\subsection{Tacit Clinical Intuition}

In contrast to this probabilistic framework stands a mode of clinical reasoning that fundamentally resists formalization. The philosophical concept of the \enquote{tacit dimension} captures the established notion that expert practitioners possess knowledge exceeding their capacity for explicit articulation; as Polanyi observes, individuals know significantly more than they can systematically formalize~\citep{polanyi2009tacit}. Formalizing this insight in the context of professional expertise, researchers note that expert performance relies upon situated, nonreflective contextual responses rather than the rigid application of formalized rules~\citep{dreyfus1986mind}. Furthermore, heuristic reasoning frequently outperforms formalized analytical models in complex, highly uncertain clinical environments~\citep{gigerenzer2007gut}.

Comparative studies of AI and human diagnostic performance reveal that this epistemological divergence is highly consequential. Physicians typically demonstrate superior performance in ambiguous, low-prevalence cases requiring nuanced contextual judgment and an understanding of ongoing patient trajectory~\citep{kumar2021ai}. Consequently, emerging ethical frameworks emphasize that clinical AI must be situated within multilevel governance structures explicitly designed to preserve the space for human interpretive and moral authority~\citep{martinho2026ethical}. This epistemological gap between probabilistic outputs and tacit judgment raises a direct regulatory question: if algorithmic architecture already constrains how physicians think, who actually governs the standard of care?

\section{\enquote{Code is Law} Applied to the Clinical Setting}

\subsection{Lessig's Framework and its Medical Translation}

Lessig's \enquote{Code is Law} thesis underscores how software architecture constrains and enables human behavior with a rigidity analogous to legal mandates. Lessig states that digital architecture sets the terms of experience, determining the baseline conditions for privacy protection and systemic control~\citep{lessig2009code}. Translated to clinical AI, this insight reveals that the fundamental architecture of a diagnostic algorithm functions as a pervasive form of de facto medical regulation. The selection of training data, the mathematical formulation of the loss function, and the specific user interface design collectively dictate how an AI system returns diagnostic differentials. This architectural structure systematically shapes the physician's cognitive attention and downstream clinical decision-making.

An inversion of Lessig's formulation, \enquote{The Law is Code,} suggests that legal principles and standards of care are increasingly translated directly into algorithmic implementations~\citep{deng2025law}. In clinical settings, this phenomenon embeds standards of care and informed consent protocols directly into software design choices. However, recent legal scholarship demonstrates that the \enquote{Code is Law} metaphor possesses inherent limitations in the context of machine learning. Unlike static legal statutes, AI systems are \emph{stochastic} (that is, their outputs contain an irreducible element of randomness, varying across runs even on identical inputs) and adaptive, meaning they update their own parameters over time. This dual character necessitates regulatory frameworks capable of responding to algorithms that autonomously evolve after initial deployment~\citep{judge2025code}. The regulation of medical AI cannot simply map Lessig's architectural logic onto a domain where the governing \enquote{code} generates dynamic, probabilistic outputs.

\subsection{Algorithmic Authority and the Standard of Care}

The traditional legal standard of care in medical malpractice faces significant disruption as AI systems become central to clinical workflows. Examinations of this \enquote{law-machine interface} demonstrate that when algorithms mediate professional medical judgment, traditional concepts of individual negligence require profound reconceptualization~\citep{yu2020artificial}. Furthermore, empirical analyses of algorithmic bias reveal that systems trained on non-representative historical datasets can systematically optimize for biased outcomes, thereby structurally disadvantaging marginalized patient populations under the guise of objective clinical recommendation~\citep{yousef2023can}.

The practical legal implications for physicians are actively manifesting in clinical practice. Scenarios are emerging wherein physicians face dual liability exposure: they may be deemed liable for incorrectly relying upon an erroneous AI recommendation, and equally liable for failing to utilize an available, highly accurate AI diagnostic tool~\citep{chew2025physicians}. Legal scholars describe this dynamic as \enquote{doctrinal collapse}: the phenomenon whereby AI technologies blur the boundaries between previously distinct legal categories, rendering established frameworks critically inadequate~\citep{solow2025ai}. To illustrate, consider that medical law traditionally separates three distinct doctrines: malpractice negligence (assessing the individual physician's deviation from standard care), vicarious liability (the hospital's responsibility for its agents), and product liability (the manufacturer's responsibility for defective software). When diagnostic reasoning is co-generated by algorithm and physician, these once-clear boundaries disintegrate. Addressing this regulatory vacuum requires policy reforms meticulously designed for adaptive technologies, grounded in continuous monitoring, and structured to establish \enquote{legal alignment} for safe AI governance~\citep{kolt2026legal}. In a practical hospital setting, legal alignment translates into institutional protocols that formally shield clinicians from liability when they critically override a flawed algorithmic recommendation, while simultaneously establishing robust procurement standards that audit AI models for equity and reliability prior to clinical deployment. Yet before designing such protocols, we must confront a deeper problem: the very outputs these algorithms produce can be confidently wrong, a failure mode that turns out to mirror a well-known weakness in human reasoning itself.

\section{The Hallucination Parallel}

The documented tendency of LLMs to generate highly plausible but factually incorrect outputs, conventionally termed \enquote{hallucination,} has emerged as a primary constraint on medical AI deployment. Research mapping the susceptibility of leading LLMs to medical misinformation finds significant variance in the capacity of different models to resist false premises embedded within clinical prompts~\citep{omar2026mapping}. This epistemic vulnerability is profoundly consequential in medical contexts, where a confident but entirely incorrect algorithmic output can directly dictate an adverse treatment decision.

However, framing hallucination purely as a localized technological defect obscures a critical structural parallel with human clinical reasoning. Confirmation bias, anchoring effects, and premature diagnostic closure are extensively documented failures of human diagnostic reasoning. These human cognitive errors share a fundamental epistemic structure with AI hallucination: both involve the generation of highly confident diagnostic assessments derived from incomplete, misunderstood, or misleading evidence~\citep{kumar2021ai}. In practice, clinicians may accept AI outputs without rigorous critical verification simply because the generated text appears structurally authoritative, thereby adopting AI-generated conclusions through a process of intuitive, unverified acceptance.

\begin{figure}[htbp]
\centering
\begin{tikzpicture}[
    auto, thick, scale=0.8, every node/.style={transform shape},
    box/.style={draw, rectangle, rounded corners, align=center, inner sep=8pt}
]

\node[box] (data) {Patient Data Streams};
\node[box, right=0.7cm of data] (ai) {AI Probabilistic Analysis};
\node[box, right=3.5cm of ai] (physician) {Physician Intuitive Assessment};
\node[box, below=0.3cm of physician] (decision) {Joint Clinical Action};

\draw[->] (data) -- (ai);
\draw[->] (ai) -- (physician) node[midway, above] {Dialectical Review};
\draw[->] (physician) -- (decision);

\draw[->] (data.north) -- ++(0, 0.5cm) -| (physician.north);

\end{tikzpicture}
\caption{Proposed Dialectical Workflow: Integrating Algorithmic Authority with Clinical Interpretive Assessment.}
\label{fig:pipeline}
\end{figure}
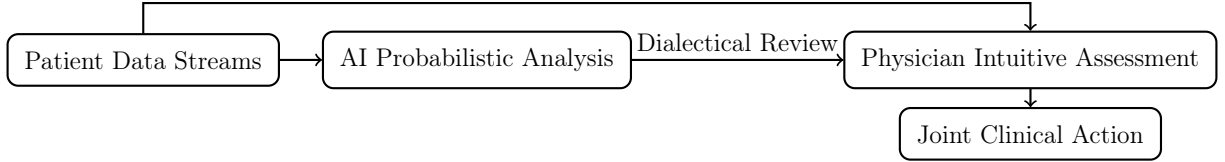

Empirical evidence confirms that even the most mathematically advanced AI systems struggle to contextualize localized clinical nuances and appropriately weight the benefits of a diagnostic test against its potential physical harms or financial costs~\citep{khera2023automation,rosbach2025automation}. This limitation highlights a sharp divergence between an algorithm's internal confidence metric and clinical diagnostic correctness~\citep{omar2025benchmarking,dratsch2023automation}. This divergence precisely mirrors the broader epistemological problem: both human cognition and algorithmic probability models suffer from structurally parallel confident errors~\citep{cajas2026beyond}. To formally address this vulnerability, Figure~\ref{fig:pipeline} outlines a proposed conceptual workflow that explicitly juxtaposes standard AI deployment against an integrated dialectical framework designed to force critical human evaluation of algorithmic outputs.

Figure~\ref{fig:pipeline} warrants careful unpacking. The workflow begins at the left with \textbf{Patient Data Streams}, representing the full range of clinical inputs: imaging, lab results, EHR fields, and genomic profiles. The first arrow feeds these data into \textbf{AI Probabilistic Analysis}, where the algorithm generates a ranked set of diagnostic hypotheses weighted by confidence scores. Crucially, the output does not proceed directly to action. Instead, the central arrow, labeled \emph{Dialectical Review}, routes the algorithmic assessment to the \textbf{Physician Intuitive Assessment} node, requiring the clinician to critically evaluate, contextualize, and either endorse or override the machine's recommendation. A separate bypass arrow runs from Patient Data Streams directly to the physician, ensuring that the clinician retains independent access to the raw clinical evidence and is never wholly dependent on the algorithm's filtered representation. Only after this structured confrontation between algorithmic output and human judgment does the workflow proceed downward to \textbf{Joint Clinical Action}, the final diagnostic or therapeutic decision attributed to the integrated AI-physician dyad. The architecture thus enforces a mandatory deliberative step that prevents the passive acceptance of machine outputs. This is precisely the cognitive safeguard needed to address the parallel vulnerabilities of hallucination and confirmation bias described above.

I believe that the policy implications of this structural parallel are substantial. If algorithmic hallucination is fundamentally analogous to human cognitive bias, regulatory frameworks must be designed to mitigate both failure modes simultaneously and interactively. Mandating formalized self-disclosure of AI-generated clinical content represents one mechanism for maintaining continuous epistemic vigilance, ensuring the provenance of clinical recommendations remains structurally transparent to the human physician~\citep{wu2026self}. Furthermore, empirical findings demonstrating that LLMs exhibit systematic demographic biases~\citep{harasta2026gender} reinforce the conclusion that algorithmic hallucination is not random noise, but rather patterned error that closely resembles heuristic biases in human judgment. Governing these parallel failure modes, however, is only possible when the underlying data infrastructure itself is trustworthy, a prerequisite that current privacy frameworks struggle to guarantee.

\section{Privacy, Data, and the Infrastructure of Trust}

The epistemological and regulatory tensions detailed above unfold within a data governance infrastructure that is fundamentally inadequate for the era of multimodal AI. Clinical AI systems demand continuous access to massive volumes of identifiable patient data for both initial training and ongoing stochastic deployment. This imperative structurally conflicts with foundational privacy protections. In the context of medical big data, this technical requirement has vastly outstripped the protective capacity of existing frameworks. HIPAA, conceptualized for an era of localized paper records, currently functions as an inadequate federal baseline. Specifically, traditional de-identification protocols (such as the HIPAA Safe Harbor method) are acutely vulnerable when applied to high-dimensional, multimodal datasets. Because advanced machine learning models can interlink diverse data points to re-identify patients, protected health information implicitly flows through opaque AI development pipelines largely unshielded by outdated privacy statutes, given that the digital transmission of health data has become entirely decoupled from the traditional therapeutic relationship~\citep{price2019privacy}.

The challenge of securing adequate data access is particularly acute in medically underserved populations. While deep learning systems theoretically offer the capacity to extend specialist-level diagnostics to remote clinical settings, this is only possible if the underlying data infrastructure supports equitable, representative deployment~\citep{solaiman2020addressing}. 

The emerging framework of AI-Supported Shared Decision-Making (AI-SDM) provides a preliminary architecture for translating abstract data governance principles into tangible clinical trust. Building directly on the \enquote{Dialectical Review} introduced in Section 4, AI-SDM functions by formally extending the critical evaluation of algorithmic outputs to patient-facing decisions. In practice, this means the physician actively translates the AI's probabilistic assessments, alongside critical caveats regarding underlying dataset limitations or demographic representation into a comprehensible clinical narrative for the patient, thereby preserving the ethical necessity of informed consent. This transparent framework establishes the AI system not as an unquestionable algorithmic oracle, but as a deliberate participant in a collaborative clinical process~\citep{as2025ai}. Nevertheless, institutional trust remains highly fragile. Rigorous transparency, technical explainability, and consistently demonstrated diagnostic accuracy have been identified as the non-negotiable prerequisites for the clinical acceptance of AI tools by healthcare workers~\citep{tun2025trust}.

The legislative response in the European Union illustrates both the theoretical promise and the practical limitations of statutory approaches to AI governance. The EU AI Act formally classifies medical AI systems as high-risk technologies, imposing rigorous statutory requirements for algorithmic transparency and mandated human oversight prior to commercial market deployment~\citep{bignami2025balancing}. Despite this robust framework, significant regulatory gaps remain unresolved, particularly concerning the dynamic deployment of continuously learning systems that alter their parameters post-deployment~\citep{vardas2025medicine}. It's evident that neither United States nor European legal frameworks currently provide comprehensive governance mechanisms for AI-powered healthcare. They specifically fail to adjudicate the complex allocation of legal responsibility when a physician and an algorithmic system collaboratively generate an adverse clinical outcome~\citep{nasir2025ethical}. The broader systemic challenge of securing independent researcher access to proprietary platform-held data further underscores the difficulty of maintaining rigorous public oversight over the corporate infrastructures dictating clinical AI~\citep{goanta2025great}. With the epistemological, regulatory, and data governance challenges now mapped, the remaining question is institutional: how should clinical practice be formally restructured to integrate algorithmic and human reasoning into a single, accountable standard of care?

\section{Toward Integration: A Dialectical Standard of Care}

\subsection{Complementarity, Not Replacement}

The term \emph{dialectical}, as used throughout this paper, denotes a structured process in which two opposing positions, here algorithmic output and physician judgment, are systematically confronted so that their synthesis yields a conclusion superior to either alone. The dialectical analysis developed here establishes that neither probabilistic algorithmic precision nor tacit human clinical authority is independently sufficient for optimal patient care. The most productive trajectory necessitates rigorous complementarity: the design of institutional architectures that explicitly position the physician as the ultimate interpretive authority responsible for contextualizing, challenging, and validating AI outputs. Building upon the paradigm of \enquote{symbiotic medicine,} clinical AI should be systematically relegated to managing foundational analytic tasks, pattern recognition, and data synthesis. Conversely, the physician must retain sovereign control over relational functions, requiring complex contextual judgment, empathetic communication, and formal moral responsibility~\citep{lee2023ai}. Evaluating the dynamics of technical protocols demonstrates that the secondary effects of these integration protocols, specifically how they subtly reshape professional clinical norms, are highly consequential~\citep{hu2025protocol}. In clinical practice, the specific design of the AI integration protocol determines whether the human-AI partnership achieves a genuine, rigorous complementarity or dangerously degrades into passive cognitive offloading by the physician. This cognitive degradation, formally characterized as \enquote{automation bias,} occurs when practitioners systematically defer to high-confidence machine outputs even when critically flawed, thereby eroding the very tacit clinical intuition the dyad is designed to preserve. In my another paper related to the AI-assisted programming has documented measurable cognitive offloading when practitioners uncritically accept machine-generated outputs~\citep{aiersilan2026vibe}; the clinical setting, where stakes are incomparably higher, is likely even more susceptible. Mitigating this risk requires interface designs that deliberately introduce cognitive friction, compelling the clinician to actively justify their concurrence or disagreement with the algorithmic recommendation.

\subsection{Specialized Medical Sub-Domains as Case Studies}

Specialized domains such as oncology and radiology provide highly instructive empirical case studies regarding effective AI integration. In the detection of prostate cancer, recent clinical evidence documents a landscape where successful technological implementations strictly position the AI system as a sophisticated triage and secondary-review layer designed to augment, rather than replace, urological expertise~\citep{rajih2025utilization}. The application of multimodal AI across oncology and cardiology demonstrates the significant potential for improved diagnostic accuracy, while simultaneously exposing the persistent, systemic challenges associated with safely integrating these tools into time-constrained clinical workflows~\citep{vela2025toward}. In adjacent diagnostic domains, AI-driven computational approaches demonstrate that machine learning models can identify and extract subtle clinical features that are mathematically complex and historically difficult to quantify manually. This capability substantially extends the diagnostic reach of the clinician. Crucially, however, it is the human clinical interpretation that synthesizes these extracted features within the holistic context of the individual patient's presentation. These case studies yield a unified structural lesson: AI integration is successful only when the algorithm is rigorously positioned as an analytical collaborator within a structured clinical workflow. It must never function as an autonomous replacement for human clinical judgment. Consequently, the physician's professional role must evolve from primary data synthesizer to the interpretive authority presiding over a vastly enriched clinical information environment.

\subsection{Regulatory and Institutional Design}

To govern this integrated framework, an active regulatory architecture must directly address the probabilistic, adaptive, and opaque nature of clinical AI. Lessig's four-modality framework of regulation, encompassing law, social norms, market forces, and technical architecture, dictates that effective governance requires synchronized intervention across all four vectors simultaneously~\citep{lessig2009code}. While the structural impacts of legal liability and technical architecture have commanded outsized regulatory attention, the dialectical standard of care critically demands equally robust interventions concerning social norms and market forces. Regarding social norms, the entrenched culture of medical training must evolve to redefine medical excellence: rather than viewing reliance on machine outputs as a degradation of skill, medical education must socialize practitioners to view the critical, structured interrogation of algorithms as a core clinical competency. Simultaneously, market forces must be aligned to sustain this dyadic relationship. This entails structuring insurance reimbursement models to adequately compensate physicians for the time-intensive process of executing the dialectical review of AI outputs, while actively combatting corporate vendor lock-in to ensure health systems retain the autonomy to audit, independently validate, and replace underperforming algorithms. This comprehensive theoretical model reinforces the demand for novel regulatory paradigms explicitly designed for systems that generate probabilistic, non-deterministic outputs. Forward-looking regulatory design must proactively anticipate aggressive technological acceleration. Because the systemic impact of clinical algorithms is profound and scalable, governance structures must be designed to be anticipatory rather than merely reactive to technological harms~\citep{neuwirth2025future}. Furthermore, the foundational ethical dimension of cyberethics mandates that any technological governance framework in healthcare must be fundamentally subordinated to the principles of human dignity, equitable access, and patient justice~\citep{spinello2010cyberethics}.

\section{Conclusion}

This paper has argued that the rapid integration of artificial intelligence into clinical medicine generates a profound epistemological tension between probabilistic algorithmic reasoning and the tacit, experiential intuition of the human physician. Rather than treating this tension as a problem to be eliminated, the analysis developed here reframes it as a productive resource: the friction between machine precision and human judgment, when institutionally structured, becomes the mechanism through which diagnostic quality is safeguarded.

Three interlocking conclusions emerge from this synthesis. First, Lessig's \enquote{Code is Law} framework reveals that algorithmic architecture is not a neutral tool but a form of de facto medical regulation. The training data, loss functions, and interface designs that constitute a clinical AI system silently constrain the physician's cognitive field, reshaping the standard of care before any formal legal rule intervenes. Second, the structural parallel between AI hallucination and human cognitive biases, including confirmation bias, anchoring, and premature closure, demonstrates that confident error is not unique to machines. This symmetry demands governance frameworks that address both failure modes within a single regulatory architecture, rather than treating algorithmic risk and human fallibility as separate problems. Third, the data governance infrastructure on which clinical AI depends remains critically inadequate: privacy statutes designed for paper records cannot secure the high-dimensional, multimodal datasets that modern AI systems require, and the resulting trust deficit threatens equitable deployment most acutely in underserved populations.

These three strands converge on a single institutional imperative: the deliberate construction of a dialectical standard of care in which the AI-physician dyad functions as the recognized unit of diagnostic responsibility. As the workflow formalized in Figure~\ref{fig:pipeline} illustrates, this standard demands mandatory structured review, a deliberative checkpoint at which the clinician must justify concurrence with, or override of, the algorithmic recommendation. Without such cognitive friction, the partnership degrades into passive automation bias, eroding the very clinical intuition it is designed to complement. 

Looking forward, two priorities are especially urgent. The first is curricular reform: medical education must socialize future practitioners to treat the critical interrogation of algorithmic outputs as a core clinical competency, not a sign of technological resistance. The second is regulatory anticipation: because clinical AI systems are stochastic and continuously adaptive, governance structures must be designed to evolve alongside the algorithms they oversee, employing continuous post-deployment auditing rather than one-time pre-market certification. The broader lesson for adjacent domains, from AI-driven body composition analysis in sarcopenia screening~\citep{aiersilan2026biotoponet,aiersilan2026literature} to autonomous vehicle safety testing, is that wherever algorithmic outputs inform high-stakes human decisions, the institutional architecture must enforce a structured dialogue between machine and practitioner.

I therefore conclude that establishing the integrated physician-AI dyad as the singular professional standard of care is not merely an institutional necessity for managing legal liability. It's a fundamental epistemological imperative: the recognition that neither silicon nor synapse alone can bear the full weight of diagnostic responsibility in the algorithmic age.


\section*{Acknowledgements}

I extend sincere thanks to Dr. C. Dianne Martin at The George Washington University for fostering the dynamic academic discussions and intellectual inquiry that inspired this manuscript.

\section*{Declaration of AI Assistance}
During the preparation of this manuscript, the authors utilized Gemini 3.1 Pro following the initial draft to assist with literature discovery and language polishing. All AI-suggested references were independently read, verified, and contextually integrated by the authors, who assume full responsibility for the manuscript's final content.


\bibliographystyle{plainnat}
\bibliography{ref}

\end{document}